\documentclass[a4paper,11pt]{article}
\usepackage{pos}
\usepackage{threeparttable}
\usepackage{hyperref}
\usepackage{amssymb}
\usepackage{color, colortbl}
\usepackage{adjustbox}
\usepackage{enumitem}

\usepackage{amssymb}
\usepackage{pifont}

\definecolor{frenchblue}{rgb}{0.0, 0.45, 0.73}
\definecolor{burgundy}{rgb}{0.5, 0.0, 0.13}
\definecolor{darkspringgreen}{rgb}{0.0, 0.65, 0.27}
\definecolor{blue}{rgb}{0.2, 0.2, 0.6}
\definecolor{indigo}{rgb}{0.0, 0.25, 0.42}
\definecolor{bluegray}{rgb}{0.4, 0.6, 0.8} 
\definecolor{lightgray}{rgb}{0.85, 0.85, 0.85}
\definecolor{gray}{rgb}{0.65, 0.65, 0.65}
\definecolor{cadetblue}{rgb}{0.37, 0.62, 0.63}
\definecolor{cambridgeblue}{rgb}{0.64, 0.76, 0.68}
\definecolor{paleaqua}{rgb}{0.74, 0.83, 0.9}
\definecolor{oldlace}{rgb}{0.99, 0.96, 0.9}
\definecolor{platinum}{rgb}{0.9, 0.89, 0.89}
\definecolor{applegreen}{rgb}{0.55, 0.71, 0.0}
\definecolor{cornsilk}{rgb}{1.0, 0.97, 0.86}
\definecolor{lavender}{rgb}{0.9, 0.9, 0.98}
\definecolor{straw}{rgb}{0.89, 0.85, 0.44}

\newcommand{\xmark}{\textcolor{burgundy}{\ding{55}}}%
\newcommand{\cmark}{\textcolor{applegreen}{\ding{51}}}%

\title{The Blazar Hadronic Code Comparison Project}
 \ShortTitle{The Blazar Hadronic Code Comparison Project}

\author*[a,b]{Matteo Cerruti}
\author[c]{Michael Kreter}
\author[d,e]{Maria Petropoulou}
\author[f]{Annika Rudolph}
\author[g]{Foteini Oikonomou}
\author[c]{Markus B{\"o}ttcher}
\author[h]{Stavros Dimitrakoudis}
\author[i]{Anton Dmytriiev}
\author[f]{Shan Gao}
\author[e]{Apostolos Mastichiadis}
\author[j]{Susumu Inoue}
\author[k,l]{Kohta Murase}
\author[m]{Anita Reimer}
\author[c]{Joshua Robinson}
\author[f]{Xavier Rodrigues}
\author[f]{Walter Winter}
\author[i]{Andreas Zech}
\author[c]{Natalia \.{Z}ywucka}

\affiliation[]{$^a$\,Universitat de Barcelona; $^b$\,Université de Paris; $^c$\,North-Western University; $^d$\,Princeton University; $^e$\,National and Kapodistrian University of Athens; $^f$\,DESY; $^g$\,Norwegian University of Science and Technology; $^h$\,University of Alberta; $^i$\,Observatoire de Paris-Meudon; $^j$\,RIKEN; $^k$\,The Pennsylvania State University; $^l$\,Yukawa Institute for Theoretical Physics; $^m$\,Leopold-Franzens-Universität Innsbruck\\}

\emailAdd{cerruti@apc.in2p3.fr}
\emailAdd{michael@kreter.org}
\emailAdd{mpetropo@phys.uoa.gr}
\emailAdd{annika.rudolph@desy.de}

\abstract{Blazar hadronic models have been developed in the past decades as an alternative to leptonic ones. In hadronic models the gamma-ray emission is associated with synchrotron emission by protons, and/or secondary leptons produced in proton-photon interactions. Together with photons, hadronic emission models predict the emission of neutrinos that are therefore the smoking gun for acceleration of relativistic hadrons in blazar jets. The simulation of proton-photon interactions and all associated radiative processes is a complex numerical task, and different approaches to the problem have been adopted in the literature. So far, no systematic comparison between the different codes has been performed, preventing a clear understanding of the underlying uncertainties in the numerical simulations. To fill this gap, we have undertaken the first comprehensive comparison of blazar hadronic codes, and the results from this effort will be presented in this contribution.\\}

\FullConference{37$^{\rm{th}}$ International Cosmic Ray Conference (ICRC 2021)\\
		July 12th -- 23rd, 2021\\
		Online -- Berlin, Germany}

\begin{document}
\maketitle

\section{Introduction}

Relativistic jets launched from accreting super-massive black holes (active galactic nuclei, AGN) radiate over the whole electromagnetic spectrum, from radio waves to TeV gamma-rays. Due to Doppler boosting, AGNs whose jet points in the direction of the observer (observationally classified as blazars) are the brightest ones. Blazars spectral energy distributions (SEDs) are composed by a non-thermal continuum that shows two well separated components: the first one peaks between infrared and X-rays, while the second one peaks in the gamma-ray band, from MeV to TeV. The origin of the first SED component is well understood as synchrotron radiation by a population of relativistic electrons/positrons in the jet. The origin of the high-energy component is still an open question. In leptonic emission scenarios, it is due to inverse Compton scattering by the same electrons/positrons that produce the first SED component. The scattering can be done on the synchrotron photons themselves (synchrotron-self-Compton process, SSC) or on external photon fields that are abundant close to the accreting black hole. In these leptonic scenarios, hadrons (protons and nuclei) might be present in the jet, but they do not contribute significantly to the emission. In hadronic scenarios, on the other hand, the second SED component is due to hadronic processes, either directly via synchrotron radiation by protons, or by secondary leptons produced in proton-photon interactions. While the two radiative models are often indistinguishable from a purely photon perspective, they can be distinguished by looking at their neutrino emission that is absent in leptonic emission processes.\\

The recent observational results from IceCube, providing evidence for neutrino emission from AGNs \cite{0506paperI,0506paperII}, have renewed interest into blazar hadronic models. With respect to leptonic models, they are more numerically challenging in particular due to the computation of proton-photon interactions, and various authors have followed different approaches for the numerical implementation. The goal of this project is to perform the first extensive comparison of the outputs from four different blazar hadronic codes codes published in the literature, estimating the level of agreement reached in the literature. In addition to the four codes, we also include in the comparison the neutrino production from simple analytical formulae. This effort is particularly relevant now that hadronic models can be tested on neutrino observations, and that predicted neutrino rates can be directly compared to results from IceCube or Antares. The spread in the outputs from the numerical codes presented here, should be regarded as an additional systematic uncertainty coming from numerical simulations.\\

The manuscript is organized as follows. In Section 2 we briefly present the four codes used in the comparison project. In Section 3 we present the result for a purely leptonic test. In Section 4 we present two different hadronic tests: a generic case in which relativistic protons interact with an arbitrary power-law photon distribution, and a more specific case where blazar $\gamma$-ray emission is powered by proton synchrotron radiation. We present our conclusions and future perspectives in Section 5.\\

\section{The numerical codes}

Simulations have been performed using four different 
hadronic radiative transfer codes: AM3, ATHE$\nu$A, Böttcher13 (B13), and LeHa-Paris. The differences among the various codes are summarized in Tables \ref{tab:processes} and \ref{tab:summary}.\\

The AM3-Code (described in \cite{Gao:2016uld}) was designed to study time-dependent multi-wavelength and multi-messenger signatures in AGNs. It computes the coupled evolution in time and energy of particle distributions for different species (photons, electrons/positrons, protons, neutrons, muons, pions and neutrinos). The processes accounted for are listed in Table~\ref{tab:processes}, secondary particles undergo the same interactions as primaries. 

ATHE$\nu$A is a time-dependent leptohadronic radiative transfer code, which was first presented in \cite{MK95}. Since then, it has been updated in various ways and evolved to its current form \citep[for details, see][]{DMPR12, PGD14}. The numerical code solves a system of coupled integro-differential equations that describe the evolution of five relativistic particle distributions within a fixed spherical volume (protons, neutrons, electrons/positrons, photons, and neutrinos). It outputs the energy spectrum of escaping radiation (and particles), after taking into account energy injection and energy losses due to various physical processes described in Table~\ref{tab:processes}. 

B13 is described in \cite{Bottcher13}; it is a steady-state code that employs an iterative scheme to evaluate equilibrium distributions of the primary electrons/positrons and protons, based on a balance of instantaneous injection (rapid acceleration) of power-law distributions of relativistic particles with radiative and adiabatic losses as well as escape. Target photon fields are the co-spatially produced primary electron synchrotron radiation and an external radiation field, approximated as isotropic in the AGN rest frame, with an arbitrary, user-defined spectrum and energy density. A semi-analytical scheme is employed to evaluate the radiative output from synchrotron-supported pair cascades following photo-hadronic processes. 

LeHa-Paris is described in \cite{Cerruti15}; it computes steady-state photon and neutrino emission from a spherical plasmoid in the jet. Primary electron and proton distributions are parameterized by broken power-law functions, while the at-equilibrium distributions of secondary particles in the plasmoid are self-consistently computed from injection and cooling terms. The pair cascades are computed iteratively generation-by-generation, under the assumption that they are never self-supported. With respect to the original version of the code, the hadronic part has been modified to accept also an arbitrary external target photon field.\\

\begin{table}
    \centering
    \begin{tabular}{c| c | c | c | c}
    \hline 
\rowcolor{indigo}   \textcolor{white}{Physical Processes} & \multicolumn{4}{c}{\textcolor{white}{Codes}} \\
    \hline 
\rowcolor{indigo}        & \textcolor{lavender}{AM3} & \textcolor{lavender}{ATHE$\nu$A} & \textcolor{lavender}{B13} & \textcolor{lavender}{LeHa-Paris}\\ 
\cellcolor{bluegray} \textcolor{white}{electron synchrotron radiation} &  \cellcolor{lightgray}   \cmark & \cellcolor{lightgray} \cmark  & \cellcolor{lightgray}   \cmark &   \cellcolor{lightgray}   \cmark   \\
\cellcolor{bluegray} \textcolor{white}{synchrotron self-absorption} & \cellcolor{lightgray}  \cmark  &  \cellcolor{lightgray} \cmark & \cellcolor{lightgray}   \cmark & \cellcolor{lightgray}  \cmark \\ 
\cellcolor{bluegray} \textcolor{white}{electron inverse Compton scattering} & \cellcolor{lightgray}   \cmark &  \cellcolor{lightgray} \cmark & \cellcolor{lightgray}  \cmark & \cellcolor{lightgray}  \cmark \\ 
\cellcolor{bluegray} \textcolor{white}{electron-positron annihilation} & \cellcolor{lightgray}  \cmark  & \cellcolor{lightgray} \cmark & \cellcolor{lightgray}  \cmark & \cellcolor{lightgray}  \xmark \\ 
\cellcolor{bluegray} \textcolor{white}{photon-photon pair production} & \cellcolor{lightgray}  \cmark  & \cellcolor{lightgray} \cmark & \cellcolor{lightgray}  \cmark & \cellcolor{lightgray}  \cmark \\ 
\cellcolor{bluegray} \textcolor{white}{triplet pair production} & \cellcolor{lightgray} \xmark &  \cellcolor{lightgray} \cmark & \cellcolor{lightgray} \xmark   & \cellcolor{lightgray} \xmark  \\ 
\cellcolor{bluegray} \textcolor{white}{proton synchrotron radiation} & \cellcolor{lightgray} \cmark  &  \cellcolor{lightgray} \cmark & \cellcolor{lightgray} \cmark   & \cellcolor{lightgray} \cmark  \\ 
\cellcolor{bluegray} \textcolor{white}{proton inverse Compton scattering} & \cellcolor{lightgray} \cmark & \cellcolor{lightgray} \xmark & \cellcolor{lightgray} \xmark & \cellcolor{lightgray}  \xmark  \\
\cellcolor{bluegray} \textcolor{white}{proton-photon pair production} &\cellcolor{lightgray} \cmark   &  \cellcolor{lightgray} \cmark & \cellcolor{lightgray}\cmark    & \cellcolor{lightgray}  \cmark  \\ 
\cellcolor{bluegray} \textcolor{white}{neutron-photon pion production} & \cellcolor{lightgray} \cmark &  \cellcolor{lightgray} \cmark & \cellcolor{lightgray} \xmark & \cellcolor{lightgray}  \xmark   \\
\cellcolor{bluegray} \textcolor{white}{kaon synchrotron radiation} & \cellcolor{lightgray} \xmark &  \cellcolor{lightgray} \cmark & \cellcolor{lightgray} \xmark & \cellcolor{lightgray}  \xmark  \\ 
\cellcolor{bluegray} \textcolor{white}{pion synchrotron radiation} &\cellcolor{lightgray} \cmark &  \cellcolor{lightgray} \cmark & \cellcolor{lightgray} \xmark & \cellcolor{lightgray}  \xmark  \\ 
\cellcolor{bluegray} \textcolor{white}{muon synchrotron radiation} & \cellcolor{lightgray} \cmark &  \cellcolor{lightgray} \cmark & \cellcolor{lightgray} \xmark & \cellcolor{lightgray}  \cmark  \\ 
    \end{tabular}
    \caption{Physical processes included in the numerical codes.}
    \label{tab:processes}
\end{table}

\begin{table}
\centering
\begin{tabular}{lcccc}
\hline 
\rowcolor{indigo}  \textcolor{white}{Features} & \multicolumn{4}{c}{ \textcolor{white}{Codes}} \\
\rowcolor{indigo}       & \textcolor{white}{AM3} & \textcolor{white}{ATHE$\nu$A} & \textcolor{white}{B13}  & \textcolor{white}{LeHA-Paris} \\ 
\hline 
\cellcolor{bluegray} \textcolor{white}{steady state} & \cellcolor{lightgray} \cmark &\cellcolor{lightgray} \cmark & \cellcolor{lightgray} \cmark & \cellcolor{lightgray}  \cmark \\
\cellcolor{bluegray} \textcolor{white}{time dependent} & \cellcolor{lightgray}\cmark & \cellcolor{lightgray}\cmark & \cellcolor{lightgray} \xmark  & \cellcolor{lightgray} \xmark \\
\cellcolor{bluegray} \textcolor{white}{linear EM cascades} &  \cellcolor{lightgray} \cmark  &  \cellcolor{lightgray} \cmark & \cellcolor{lightgray} \cmark & \cellcolor{lightgray}  \cmark \\
\cellcolor{bluegray} \textcolor{white}{non-linear EM cascades} &  \cellcolor{lightgray}  \cmark &   \cellcolor{lightgray}\cmark &   \cellcolor{lightgray}\xmark &  \cellcolor{lightgray}\xmark \\
\rowcolor{indigo} \textcolor{white}{Implementation} & \multicolumn{4}{c}{}\\ 
\hline
\cellcolor{bluegray} \textcolor{white}{$p \gamma \pi$ processes} &  \cellcolor{lightgray} following \cite{Hummer:2010vx} &  \cellcolor{lightgray} tabulated {\sc sophia} \cite{Mucke00} & \cellcolor{lightgray} following \cite{Kelner_2008}  &  \cellcolor{lightgray} running {\sc sophia} \cite{Mucke00}\\
\cellcolor{bluegray} \textcolor{white}{$p \gamma e$ processes} &  \cellcolor{lightgray} following \cite{Kelner_2008}  &  \cellcolor{lightgray} tabulated from \cite{PJ96} &  \cellcolor{lightgray}  following \cite{Kelner_2008} &  \cellcolor{lightgray} following \cite{Kelner_2008} \\
\end{tabular}
\caption{Main features of numerical codes and implementation of hadronic processes.}
\label{tab:summary}
\end{table}
\section{Semi-analytic approximation}

We also compare the neutrino fluxes obtained with the four numerical codes to the neutrino fluxes obtained 
with a simple semi-analytic approach. This approach is also standard in the literature and we are keen to
test the range of applicability of this method. We define the photomeson production timescale for protons with Lorentz factor $\gamma_p$ as~\cite{1979ApJ...228..919S},
\begin{equation} 
t_{\rm p\gamma}^{-1} (\gamma_p) = \frac{c}{2 \gamma_p^2}
\int^{\infty}_{\epsilon_{\rm th}} {\rm d}
\epsilon_{\gamma} \sigma_{\rm p \gamma}(\epsilon_{\gamma}) \kappa_{\rm p \gamma}(\epsilon_{\gamma})
\epsilon_{\gamma}
\int^{\infty}_{\epsilon_{\gamma}/(2\gamma_p)} {\rm d}
\epsilon \epsilon^{-2}
  n_{\rm ph},
\label{eq:pgammaRate}
\end{equation}
\noindent where, $\epsilon_{\gamma}$ is the photon energy in the proton rest frame, 
$\epsilon_{\rm th} \sim 145$~MeV is the threshold energy for pion production, and
$\sigma_{\rm p \gamma}$ and $\kappa_{\rm p \gamma}$ are the cross-section and
inelasticity of photomeson interactions, respectively. 
We use the parametrisations of~\cite{Murase:2005hy} for $\sigma_{\rm  p \gamma}$ and
$\kappa_{\rm  p \gamma}$. The quantity
$n_{\rm ph}$ is the spectral number density of target photons with energy $\epsilon$. 
The fraction of energy converted to pions is estimated as,
$f_{\rm p \gamma} \equiv t_{\rm cool} / t_{\rm p\gamma}$, where $t_{\rm cool}$ is the 
proton energy loss cooling time, defined as
\begin{equation} 
t_{\rm cool}^{-1} \equiv t_{\rm cross}^{-1} + t_{p,{\rm syn}}^{-1} + t_{\rm p \gamma}^{-1},
\end{equation}
where the synchrotron cooling time for protons with energy $\varepsilon_p$ 
in a magnetic field with strength $B$
is given by, $t_{p,{\rm syn}} = 6 \pi m_p^4c^3/(m_e^2 \sigma_T B^2
\varepsilon_p)$.
Here, $m_p$ and $m_e$ are the proton and electron mass respectively, $c$ the speed of light, and $\sigma_T$ the Thomson cross section. The crossing time, $t_{\rm cross} = r_{b}/c$, approximates the adiabatic energy loss rate. The all flavour neutrino luminosity per logarithmic energy is estimated as,
$$
\varepsilon_{\nu}L_{\varepsilon_{\nu}} \approx \frac{3}{8} f_{p\gamma}(\varepsilon_{p}) \varepsilon_{p}L_{\varepsilon_{p}}
$$
where $\varepsilon_{p}L_{\varepsilon_{p}}$ is the injected proton luminosity per logarithmic energy
and the neutrinos are assumed to be produced with energy $\varepsilon_{\nu} \sim 0.05 \varepsilon_p$. \\

\section{Leptonic comparisons}

Our first test is a simple Synchrotron-Self-Compton (SSC) scenario with inverse Compton scatterings occurring in Thomson regime. 
The injected electron distribution of power law $\alpha_e = 1.9$ extends between $\gamma_{e, min} = 1$ and $\gamma_{e, max} = 10^{4}$ (with an exponential cutoff) with an injection compactness of $\log_{10}(l_{e, inj}) = - 4.47$. To reduce the effect of cooling on the lepton distribution, we choose a magnetic field of $B=0.01$~G; the radius of the spherical emission region is $R = 10^{15}$~cm, which allows to calcualte the escape time for all particle species as $t_{esc} = R/ \mathrm{c}$. For better comparability of the pure SSC process, photon-photon annihilation was neglected.

\begin{figure}[th]
\begin{center}
\includegraphics[width= 0.6\textwidth]{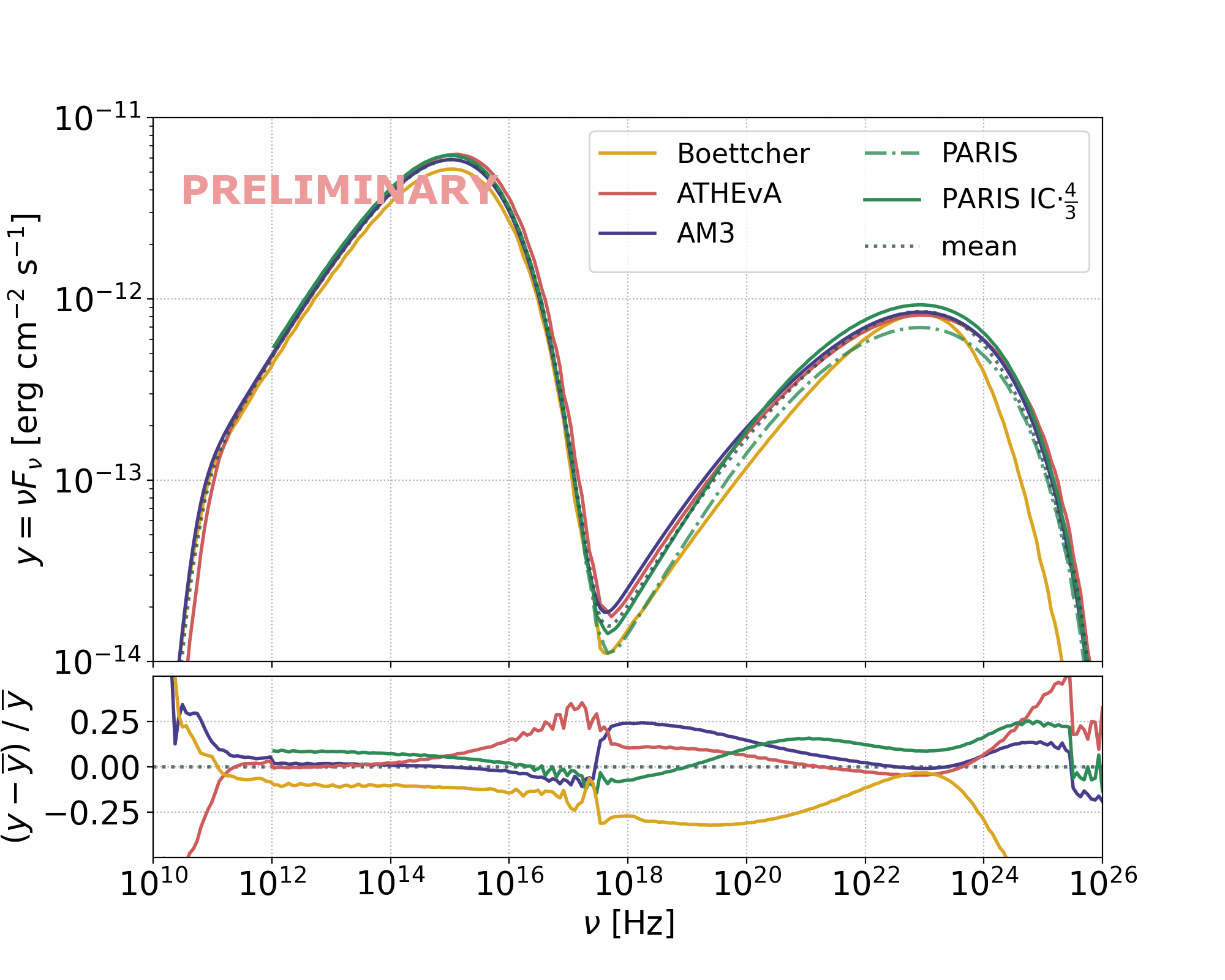}\\
\caption{Synchrotron-self-Compton test: Comparison of the observed steady-state spectra. The mean is calculated using the 'Paris IC $\cdot \frac{3}{4}$' - result that corrects for the geometrical averaging factor which is not used by the other codes.}
\label{fig:ssc}
\end{center}
\end{figure}

A comparison of the observed steady-state spectra for the different Codes is shown in Fig.~\ref{fig:ssc}, where we assumed a Doppler factor of $\delta = 30$ and a redshift of $ z = 0.01$. 

Overall we find good agreement between the different codes for the synchrotron component. To account for spatial averaging in a sphere, the LeHa-Paris multiplies the synchrotron target photons with a factor of $3/4$ for the inverse Compton scatterings. If we correct for this effect (solid green line in Fig.~\ref{fig:ssc}), the results differ less than 20 \% at the inverse Compton peak.
 
\section{Hadronic comparisons}

\begin{figure}[th]
\begin{center}
\includegraphics[width= 0.9\textwidth]{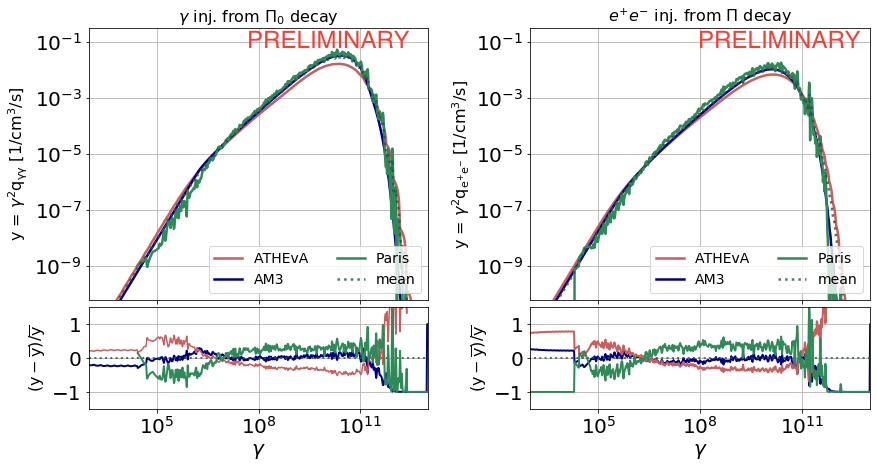}\\
\includegraphics[width= 0.45\textwidth]{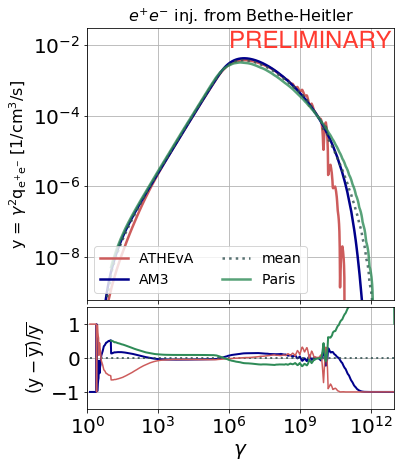}
\includegraphics[width= 0.45\textwidth]{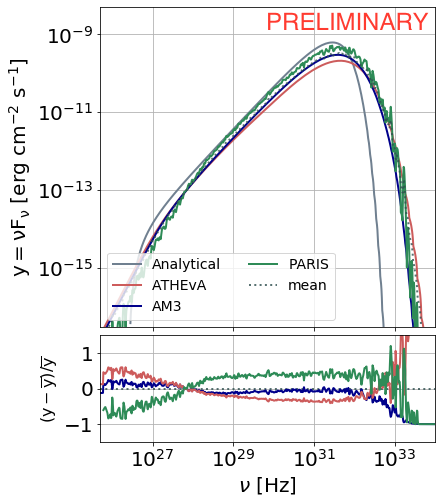}
\caption{Hadronic test using a power-law target photon field for proton-photon interactions. Top left: spectra of secondary photons from neutral pion decay; top right: spectra of electrons/positrons from charged pion decay;  bottom left: spectra of electrons/positrons from Bethe-Heitler pair-production; bottom right: neutrino spectra in the observer's frame.}
\label{fig:powerlaw}
\end{center}
\end{figure}

As first hadronic test, we compare the results from the interaction of protons in the plasmoid with an external power-law distribution of photons. The primary proton population is parameterized by a power-law function with index $\alpha_p = 1.9$ and exponential cut-off at $\gamma_{p, max} = 10^8$. The normalization of the proton injection is $\log_{10}(l_{p, inj}) = - 4$. The photon field is parameterized by a power-law function with index $\alpha_{ph} = 2$ between the minimum and maximum reduced energies ($E/mc^2$) $\epsilon_{min} = 10^{-6}$ and $\epsilon_{Max} = 0.1$. The remaining free parameters of the emitting region are $R = 10^{15}$ cm, $B = 10$ G and $\delta = 30$, and a redshift of $ z = 0.01$. 
In Figure \ref{fig:powerlaw} we show four distinct hadronic processes: injection of photons from $\pi_0$ decay, injection of $e^\pm$ from $\pi^\pm$ decay, injection of $e^\pm$ from Bethe-Heitler, and the neutrino spectra in the observer's frame (all flavours).\\

\begin{figure}[th]
\begin{center}
\includegraphics[width= 0.9\textwidth]{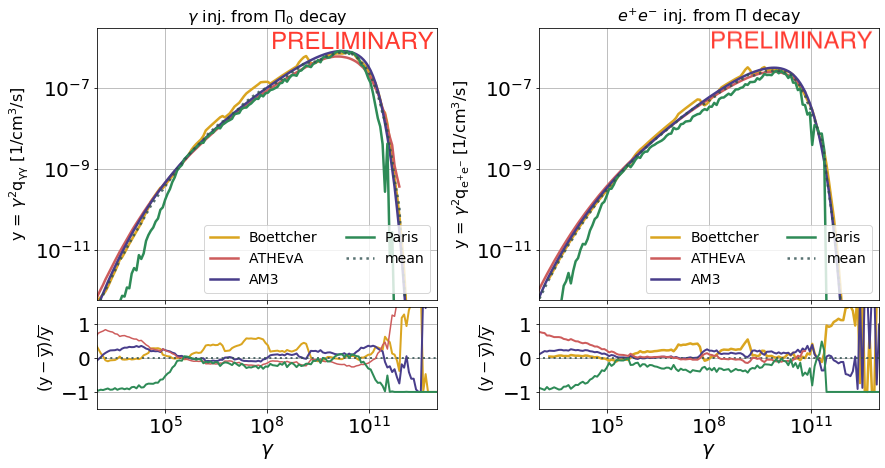}\\
\includegraphics[width= 0.45\textwidth]{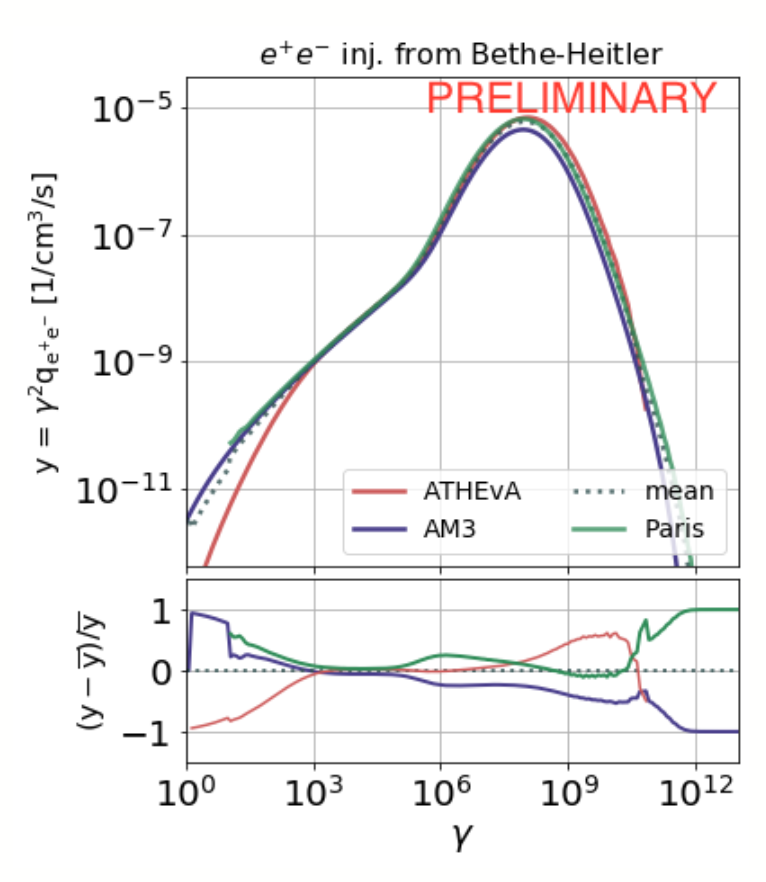}
\includegraphics[width= 0.45\textwidth]{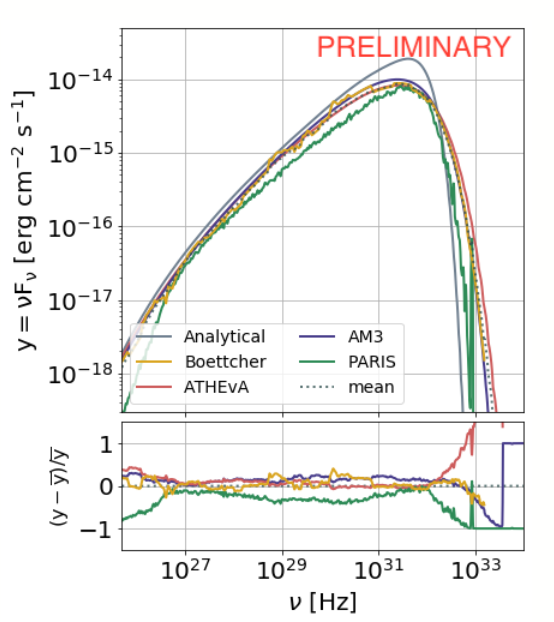}
\caption{Proton synchrotron test, see caption of Figure \ref{fig:powerlaw} for details.}
\label{fig:psyn}
\end{center}
\end{figure}

As second hadronic test, we compute the photon and neutrino emission from a more realistic blazar proton-synchrotron model. In this case there are no external photon fields, and the target photon field for p-$\gamma$ interactions is synchrotron emission by primary electrons in the jet. The primary proton population is parameterized by a power-law function with index $\alpha_p = 1.9$ and exponential cut-off at $\gamma_{p, max} = 10^8$, with normalization $\log_{10}(l_{e, inj}) = - 4.93$. The electron population is parameterized by a power-law function with index $\alpha_e = 1.9$ and exponential cut-off at $\gamma_{e, max} = 10^3$, with normalization $\log_{10}(l_{e, inj}) = - 7.47$ The other free parameters of the emitting region are identical to the first hadronic test. In Figure \ref{fig:psyn} we show the results of our simulations, in the same way as for the power-law photon-field case.\\

\section{Conclusions and perspectives}

We have performed the first extensive comparison of outputs from blazar hadronic codes. The preliminary results shown here indicate that there is a good agreement in terms of spectral shapes, besides some distortions at low/high-energy cut-offs. We do see a spread in normalizations, both for photo-meson interactions and Bethe-Heitler pair production. We preliminary estimate this spread at the level of 30-40$\%$.\\

The results shown here represent only a part of the tests we performed. We plan to publish comparison tests for external-inverse-Compton processes, hadronic interactions with mono-energetic protons, high-opacity scenario with significant proton cooling, and time-dependent signatures.
All outputs will be openly released as online material once the paper will be published, to serve as benchmark for future code developments. Feel free to contact us already if you would like to get earlier access to the preliminary curves.

\bibliographystyle{ICRC}
\bibliography{references}

\providecommand{\href}[2]{#2}\begingroup\raggedright\begin{thebibliography}{10}

\bibitem{0506paperI}
{IceCube Collaboration}, {Fermi-LAT Collaboration}, {MAGIC Collaboration},
  {AGILE Team}, {ASAS-SN Team}, {HAWC Collaboration}, {H.~E.~S.~S.
  Collaboration}, {INTEGRAL Team}, {VERITAS Collaboration}, and {VLA/B Team}
  \href{http://dx.doi.org/10.1126/science.aat1378}{{\em Science} {\bfseries
  361} no.~6398, (July, 2018) eaat1378}.

\bibitem{0506paperII}
{IceCube Collaboration} \href{http://dx.doi.org/10.1126/science.aat2890}{{\em
  Science} {\bfseries 361} no.~6398, (July, 2018) 147--151}.

\bibitem{Gao:2016uld}
S.~Gao, M.~Pohl, and W.~Winter
  \href{http://dx.doi.org/10.3847/1538-4357/aa7754}{{\em Astrophys. J.}
  {\bfseries 843} no.~2, (2017) 109}.

\bibitem{MK95}
A.~{Mastichiadis} and J.~G. {Kirk} {\em A\&A} {\bfseries 295} (Mar., 1995) 613.

\bibitem{DMPR12}
S.~{Dimitrakoudis}, A.~{Mastichiadis}, R.~J. {Protheroe}, and A.~{Reimer}
  \href{http://dx.doi.org/10.1051/0004-6361/201219770}{{\em A\&A} {\bfseries
  546} (Oct., 2012) A120}.

\bibitem{PGD14}
M.~{Petropoulou}, D.~{Giannios}, and S.~{Dimitrakoudis}
  \href{http://dx.doi.org/10.1093/mnras/stu1757}{{\em MNRAS} {\bfseries 445}
  no.~1, (Nov., 2014) 570--580}.

\bibitem{Bottcher13}
M.~{B{\"o}ttcher}, A.~{Reimer}, K.~{Sweeney}, and A.~{Prakash}
  \href{http://dx.doi.org/10.1088/0004-637X/768/1/54}{{\em ApJ} {\bfseries 768}
  no.~1, (May, 2013) 54}.

\bibitem{Cerruti15}
M.~{Cerruti}, A.~{Zech}, C.~{Boisson}, and S.~{Inoue}
  \href{http://dx.doi.org/10.1093/mnras/stu2691}{{\em MNRAS} {\bfseries 448}
  no.~1, (Mar., 2015) 910--927}.

\bibitem{Hummer:2010vx}
S.~Hummer, M.~Ruger, F.~Spanier, and W.~Winter
  \href{http://dx.doi.org/10.1088/0004-637X/721/1/630}{{\em Astrophys. J.}
  {\bfseries 721} (2010) 630--652}.

\bibitem{Mucke00}
A.~{M{\"u}cke}, R.~{Engel}, J.~P. {Rachen}, R.~J. {Protheroe}, and T.~{Stanev}
  \href{http://dx.doi.org/10.1016/S0010-4655(99)00446-4}{{\em Computer Physics
  Communications} {\bfseries 124} no.~2-3, (Feb., 2000) 290--314}.

\bibitem{Kelner_2008}
S.~R. {Kelner} and F.~A. {Aharonian}
  \href{http://dx.doi.org/10.1103/PhysRevD.78.034013}{{\em PrD} {\bfseries 78}
  no.~3, (Aug., 2008) 034013}.

\bibitem{PJ96}
R.~J. {Protheroe} and P.~A. {Johnson}
  \href{http://dx.doi.org/10.1016/0927-6505(95)00039-9}{{\em Astroparticle
  Physics} {\bfseries 4} no.~3, (Feb., 1996) 253--269}.

\bibitem{1979ApJ...228..919S}
F.~W. {Stecker} \href{http://dx.doi.org/10.1086/156919}{{\em ApJ} {\bfseries
  228} (Mar., 1979) 919--927}.

\bibitem{Murase:2005hy}
K.~Murase and S.~Nagataki
  \href{http://dx.doi.org/10.1103/PhysRevD.73.063002}{{\em Phys. Rev.}
  {\bfseries D73} (2006) 063002}.

\end{thebibliography}\endgroup

%
%
%

\end{document}